\documentclass[11pt]{amsart} 
\usepackage{amscd,amssymb,amsxtra}
\usepackage{mathrsfs}  % \mathrsfs{A}
\DeclareSymbolFontAlphabet{\mathrsfs}{rsfs}
\usepackage[mathscr]{eucal} % \mathscr{A} 
\usepackage{comment}
\usepackage{color}
\usepackage{enumitem} 
\usepackage{bbm}
\usepackage[colorlinks,backref=page,citecolor=refkey]{hyperref}
\usepackage{aliascnt}
\usepackage[nomath]{lmodern}

\usepackage{marginnote}

\makeatletter
\let\@secnumfont\bfseries
\def\section{\@startsection{section}{1}%
  \z@{4\linespacing\@plus\linespacing}{\linespacing}%
  {\bfseries\centering}}
\def\introsection{\@startsection{section}{1}%
  \z@{3\linespacing\@plus\linespacing}{\linespacing}%
  {\bfseries\centering}}
\def\subsection{\@startsection{subsection}{2}%
   \z@{1.25\linespacing\@plus.7\linespacing}{.5\linespacing}%
   {\normalfont\bfseries}}
\def\subsectionsinline{\def\subsection{\@startsection{subsection}{2}%
  \z@{1\linespacing\@plus.7\linespacing}{-.5em}%
  {\normalfont\bfseries}}}

\makeatother

\numberwithin{equation}{section}

\newcommand{\mynewtheorem}[2]{
  \newaliascnt{#1}{equation}
  \newtheorem{#1}[#1]{#2}
  \aliascntresetthe{#1}
  \expandafter\def\csname #1autorefname\endcsname{#2}
}

\theoremstyle{definition}
\mynewtheorem{definition}{Definition}
\mynewtheorem{example}{Example}
\mynewtheorem{problem}{\color{blue}Problem}
\mynewtheorem{probsec}{\color{blue}Problem}
\mynewtheorem{exercise}{Exercise}
\mynewtheorem{question}{\color{blue}Question}
\mynewtheorem{project}{\color{blue}Project}
\mynewtheorem{construction}{Construction}
\mynewtheorem{task}{\color{blue}Task}
\mynewtheorem{otask}{\color{green}Obsolete Task}
\mynewtheorem{notation}{Notation}

\newtheorem*{definition*}{Definition}
\newtheorem*{example*}{Example}
\newtheorem*{problem*}{\color{blue}Problem}
\newtheorem*{probsec*}{\color{blue}Problem}
\newtheorem*{exercise*}{Exercise}
\newtheorem*{question*}{\color{blue}Question}
\newtheorem*{project*}{\color{blue}Project}
\newtheorem*{construction*}{Construction}
\newtheorem*{notation*}{Notation}

\theoremstyle{remark}
\mynewtheorem{note}{Note}
\mynewtheorem{remark}{Remark}
\mynewtheorem{data}{Data}

\newtheorem*{note*}{Note}
\newtheorem*{remark*}{Remark}
\newtheorem*{data*}{Data}

\theoremstyle{plain}
\mynewtheorem{theorem}{Theorem}
\mynewtheorem{corollary}{Corollary}
\mynewtheorem{lemma}{Lemma}
\mynewtheorem{proposition}{Proposition}
\mynewtheorem{conjecture}{Conjecture}
\mynewtheorem{claim}{Claim}
\mynewtheorem{proposal}{Proposal}
\mynewtheorem{conclusion}{Conclusion}
\mynewtheorem{hypothesis}{Hypothesis}
\mynewtheorem{assumption}{Assumption}

\newtheorem*{theorem*}{Theorem}
\newtheorem*{corollary*}{Corollary}
\newtheorem*{lemma*}{Lemma}
\newtheorem*{proposition*}{Proposition}
\newtheorem*{conjecture*}{Conjecture}
\newtheorem*{claim*}{Claim}
\newtheorem*{proposal*}{Proposal}
\newtheorem*{conclusion*}{Conclusion}
\newtheorem*{hypothesis*}{Hypothesis}
\newtheorem*{assumption*}{Assumption}

\newenvironment{proof*}[1][\proofname]{
  \begin{proof}[#1]}{  
\end{proof}}

\definecolor{refkey}{rgb}{0,.6,.4}

\renewcommand{\:}{\colon}
\newcommand{\CC}{{\mathbb C}}

\DeclareMathOperator{\End}{End}

\DeclareMathOperator{\Hom}{Hom}
\DeclareMathOperator{\id}{id}

\DeclareMathOperator{\Map}{Map}

\newcommand{\RP}{{\mathbb R\mathbb P}}
\newcommand{\RR}{{\mathbb R}}

\newcommand{\ZZ}{{\mathbb Z}}

\newcommand{\chiup}{\raise.5ex\hbox{$\chi$}}

\newcommand{\inv}{^{-1}}
\DeclareRobustCommand{\mstrut}{^{\vphantom{1*\prime y\vee M}}}

\newcommand{\res}[1]{\negmedspace\bigm|\mstrut_{#1}}

\newcommand{\temsquare}{\raise3.5pt\hbox{\boxed{ }}}

\newcommand{\hneg}{\mkern-.5\thinmuskip}
 
\DeclareFontFamily{U}{mathx}{}
\DeclareFontShape{U}{mathx}{m}{n}{<-> mathx10}{}
\DeclareSymbolFont{mathx}{U}{mathx}{m}{n}
\DeclareMathAccent{\widehat}{0}{mathx}{"70}
\DeclareMathAccent{\widecheck}{0}{mathx}{"71}
 
\DeclareMathSymbol{\bigtimes}{1}{mathx}{"91}

\DeclareMathOperator{\SO}{SO}

\DeclareMathOperator{\SU}{SU}

\DeclareMathOperator{\SL}{SL}
\DeclareMathOperator{\PSL}{PSL}

\usepackage[all,2cell]{xy}
\usepackage{graphicx}\usepackage{epsf}

\definecolor{refkey}{rgb}{0,.8,.2}\definecolor{labelkey}{rgb}{1,0,0} 

\newcommand{\bmuu}{\mbox{$\raisebox{-.07em}{\rotatebox{9.9}
  {\tiny {\bf /}
  }}\hspace{-0.53em}\mu\hspace{-0.88em}\raisebox{-0.98ex}{\scalebox{2} 
  {$\color{white}\phantom{.}$}}\hspace{-0.416em}\raisebox{+0.88ex}
  {$\color{white}\phantom{.}$}\hspace{0.46em}$}} 
\newcommand{\bmut}{\bmu 2}
\newcommand{\bmu}[1]{\bmuu _{#1}}

\DeclareMathOperator{\Cat}{Cat}

\DeclareMathOperator{\Fun}{Fun}
\DeclareMathOperator{\Ind}{Ind}

\DeclareMathOperator{\Vect}{Vect}

\newcommand{\Aq}{A',q}

\newcommand{\BnA}[1]{B^{#1}\hneg A}

\newcommand{\BtAp}{B^2\!A'}
\newcommand{\BtA}{B^2\!A}
\newcommand{\Cx}{\CC^{\times}}
\newcommand{\GA}{\CC[G]}

\newcommand{\RAq}{R_{A',q}}

\newcommand{\Usl}{\mathcal{U}(\sltR)}
\newcommand{\VAq}{\sV_{\Aq}}
\newcommand{\VV}{\mathbb{V}}

\newcommand{\Xm}[1]{\sX^{#1}}

\newcommand{\bH}{\overline{H}}

\newcommand{\bone}{\mathbbm{1}}
\newcommand{\bs}{\!\!\!\bigm/\!\!\sigma }
\newcommand{\dual}{^\vee}
\newcommand{\eA}{\epsilon \mstrut _{\hneg A}}

\newcommand{\sC}{\mathscr{C}}

\newcommand{\sG}{\mathscr{G}}

\newcommand{\sK}{\mathcal{K}}
\newcommand{\sL}{\mathcal{L}}
\newcommand{\sM}{\mathscr{M}}

\newcommand{\sV}{\mathcal{V}}

\newcommand{\sX}{\mathscr{X}}
\newcommand{\sY}{\mathscr{Y}}

\newcommand{\scaP}{\mathscr{P}}

\newcommand{\sltR}{\mathfrak{s}\mathfrak{l}_2\RR}

\newcommand{\sr}{(\sigma ,\rho )}
\newcommand{\tFt}{(\tF,\theta )}
\newcommand{\tF}{\widetilde{F}}

\renewcommand{\flat}{\mstrut _{\textnormal{flat}}}
\newcommand{\gpd}{/\!/}

\DeclareTextFontCommand{\textbfit}{\bfseriesitshape}

\begin{document}

\abovedisplayskip18pt plus4.5pt minus9pt
\belowdisplayskip \abovedisplayskip
\abovedisplayshortskip0pt plus4.5pt
\belowdisplayshortskip10.5pt plus4.5pt minus6pt
\baselineskip=15 truept
\marginparwidth=55pt

\makeatletter
\DeclareRobustCommand\bfseriesitshape{%
  \not@math@alphabet\itshapebfseries\relax
  \fontseries\bfdefault
  \fontshape\itdefault
  \selectfont
}
\renewcommand{\tocsection}[3]{%
  \indentlabel{\@ifempty{#2}{\hskip1.5em}{\ignorespaces#1 #2.\;\;}}#3}
\renewcommand{\tocsubsection}[3]{%
  \indentlabel{\@ifempty{#2}{\hskip 2.5em}{\hskip 2.5em\ignorespaces#1%
    #2.\;\;}}#3} 
\renewcommand{\tocsubsubsection}[3]{%
  \indentlabel{\@ifempty{#2}{\hskip 5.5em}{\hskip 5.5em\ignorespaces#1%
    #2.\;\;}}#3} 
%  \indentlabel{\hskip 4em#3}}
\renewcommand\subsubsection{\@startsection{subsubsection}{3}%
  \z@{.5\linespacing\@plus.7\linespacing}{-.5em}%
  {\normalfont\bfseriesitshape}}
\def\@makefnmark{%
  \leavevmode
  \raise.9ex\hbox{\fontsize\sf@size\z@\normalfont\tiny\@thefnmark}} 
\def\multfoot{\textsuperscript{\tiny\color{red},}}
\def\footref#1{$\textsuperscript{\tiny\ref{#1}}$}
%             \footnote{Note\label{SYM}} ... in footnote~\footref{SYM}
\makeatother

\newcommand{\lact}{\;\reflectbox{\rotatebox[origin=c]{+90}{$\circlearrowleft$}}\;} 
\newcommand{\ract}{\;\rotatebox[origin=c]{+90}{$\circlearrowleft$}\;} 
\newcommand{\squig}{{\scriptstyle\sim\mkern-3.9mu}}
\newcommand{\lsquigend}{{\scriptstyle\lhd\mkern-3mu}}
\newcommand{\rsquigend}{{\scriptstyle\rule{.1ex}{0ex}\rhd}}
\newcounter{sqindex}
\newcommand\squigs[1]{%
  \setcounter{sqindex}{0}%
  \whiledo {\value{sqindex}< #1}{\addtocounter{sqindex}{1}\squig}%
}
\newcommand\rsquigarrow[2]{%
  \mathbin{\stackon[2pt]{\squigs{#2}\rsquigend}{\scriptscriptstyle\text{#1\,}}}%
}
\newcommand\lsquigarrow[2]{%
  \mathbin{\stackon[2pt]{\lsquigend\squigs{#2}}{\scriptscriptstyle\text{\,#1}}}%
}

\setcounter{tocdepth}{2}

\renewcommand{\theequation}{\arabic{equation}}
\renewcommand{\theremark}{\arabic{remark}} 
\renewcommand{\thedefinition}{\arabic{definition}} 
\renewcommand{\theexample}{\arabic{example}} 

 \title[Topological symmetry in QFT]{Introduction to topological symmetry in QFT}
 \author[D. S. Freed]{Daniel S.~Freed}
 \thanks{This material is based upon work supported by the National Science
Foundation under Grant Number DMS-2005286 and by the Simons Foundation Award
888988 as part of the Simons Collaboration on Global Categorical Symmetries.
It is also based upon work supported by the National Science Foundation under
Grant No. 1440140, while the author was in residence at the Mathematical
Sciences Research Institute in Berkeley, California, during the fall semester
of~2022.}
 \address{Department of Mathematics \\ University of Texas \\ Austin, TX
78712} 
 \email{dafr@math.utexas.edu}
% \dedicatory{}
 \date{July 5, 2023}
 \begin{abstract} 
 This brief note publicizes the quantum framework for symmetry that is
developed in our joint paper~\cite{FMT} with Greg Moore and Constantin
Teleman.  We include additional motivation and an application to a selection
rule for line defects in 4-dimensional gauge theories.
 \end{abstract}
\maketitle

This note\footnote{which will appear in the conference proceedings of String
Math 2022} is a preview (appearing in arrears) of joint work with Greg Moore
and Constantin Teleman~\cite{FMT}, to which we refer for further exposition,
further development, more examples, and for extensive references to the
literature, since we omit citations here.

As a first introduction to our framework for topological symmetries, consider
the simpler setting of isometries in Riemannian geometry.  Let $M$~be a
Riemannian manifold.  In differential geometry we define the notions of (i)~a
Lie group~$G$, and (ii)~the action of~$G$ on~$M$ by isometries.  These are both
fairly straightforward definitions.  What is more difficult (Myers-Steenrod
theorem) is to prove that the group of \emph{all} isometries of~$M$ has a
canonical Lie group structure.  Our main definitions for topological symmetries
in quantum field theory are analogous to~(i) and~(ii); we do not attempt to
give structure to the collection of all topological symmetries of a quantum
field theory.\footnote{\label{confess}There are well-developed mathematical
foundations for \emph{topological} field theory, but the geometric development
of general quantum field theories is at a much more nascent phase.  So while we
treat topological field theories rigorously, we take a more heuristic approach
to nontopological quantum field theories.  Therefore, we do not even attempt a
definition of topological symmetries in a nontopological quantum field theory.}

Here are the main points:

 \begin{enumerate}[label=\textnormal{(\arabic*)}]

 \item We define separately (i)~abstract symmetry data, and (ii)~a concrete
realization of abstract symmetry data on a quantum field theory.  The
data~(i) is a \emph{quiche}\footnote{This strange terminology for an object
that encodes abstract \emph{qu}antum symmetries evokes a pair~$\sr$ in which
$\rho $~is the crust and $\sigma $~the filling; imagine the sandwich on the
left of \autoref{fig:3} made open-faced by omitting the right ``piece of
bread''~$\tF$.  One can (and one did!) point out that a \emph{qu}iche is
\emph{not} an open-faced sandwich.  (Google `tartine'.)}~$\sr$ in which
$\sigma $~is a \emph{topological} field theory and $\rho $~is a
\emph{topological} boundary theory.  The data of the action~(ii) of~$\sr$ on
a quantum field theory~$F$ belongs to nontopological field theory if $F$~is
not topological.

 \item Topological symmetry is treated as a \emph{quantum} object: the
pair~$\sr$.  This goes beyond the usual coupling of a theory to a background
gauge field, which treats symmetry in a \emph{classical} context.  We
illustrate the power of this perspective by means of an example at the end of
this note.

 \item Our definitions are inspired by an analogy that we explicate below:  
  \begin{equation}\label{eq:1}
     \begin{aligned} \text{the pair~$(A,R)$ of an algebra~$A$ and a right
      module~$R$} &\mathbin{\lsquigend\squig\squig\squig\rsquigend} \text{a quiche
     $\sr$}\\ \text{element 
      of  $A$} &\mathbin{\lsquigend\squig\squig\squig\rsquigend}
     \textnormal{defect in~$\sr$}\end{aligned} 
  \end{equation}

 \end{enumerate}

\noindent
 Our framework applies to internal topological symmetries of quantum field
theories.  It includes homotopical symmetries: higher groups, 2-groups, etc.

To reinforce the power of separating abstract symmetry from its concrete
action, we begin with an illustration from representation theory.  Then we
recount the key definitions from~\cite{FMT}.  We end with an application of
our framework to 4-dimensional gauge theory, or more to properly
4-dimensional quantum field theories with a certain symmetry: the derivation
of a selection rule for line defects.
 
I warmly thank Greg Moore and Constantin Teleman for their collaboration and
their feedback on this note.

  \subsection*{Motivation: Computations in representations of $\sltR$}

\subsubsection*{2-dimensional representation} Set  
  \begin{equation}\label{eq:3}
     h= \begin{pmatrix} 1 & 0 \\ 0 & -1 \\ \end{pmatrix} \qquad
     e= \begin{pmatrix} 0 & 1 \\ 0 & 0 \\ \end{pmatrix} \qquad
     f= \begin{pmatrix} 0 & 0 \\ 1 & 0 \\ \end{pmatrix}
  \end{equation}
These matrices form a basis of the Lie algebra~$\sltR$.  Straightforward
addition and multiplication of $2\times 2$ matrices verifies the identity 
  \begin{equation}\label{eq:2}
     \frac 12 h^2 + ef + fe = \frac 12h^2 + h + 2fe 
  \end{equation}
In fact, both sides equal the scalar matrix $\left(\begin{smallmatrix} 
3/2&0\\[2pt]0& 3/2 \end{smallmatrix}\right)$. 
 
\subsubsection*{3-dimensional representation} Here the matrices that
represent the elements~\eqref{eq:3} of~$\sltR$ are
  \begin{equation}\label{eq:4}
     h'= \begin{pmatrix} 2 & 0 & 0 \\ 0 & 0 & 0 \\ 0 & 0 & -2 \\
     \end{pmatrix} \qquad e'= \begin{pmatrix} 0 & 1 & 0 \\ 0 & 0 & 2 \\
     0 & 0 & 0 \\ \end{pmatrix} \qquad f'= \begin{pmatrix} 0 & 0 & 0 \\
     2 & 0 & 0 \\ 0 & 1 & 0 \\ \end{pmatrix} 
  \end{equation}
One can compute---perhaps not quite as easily---that the
identity~\eqref{eq:2} holds:
  \begin{equation}\label{eq:5}
     \frac 12 (h')^2 + e'f' + f'e' = \frac 12(h')^2 + h' + 2f'e' 
  \end{equation}
Again both sides are scalar matrices, namely 4~times the identity matrix.
 
\subsubsection*{Infinite dimensional representations} The Lie group
$\SL_2\!\RR$ acts on the projective line~$\RP^1$ as fractional linear
transformations, effectively through its quotient $\PSL_2\!\RR$.  For each
complex number $\lambda \in \CC$, there is an induced action on $\lambda
$-differentials $\phi (x)(dx)^\lambda $.  The infinitesimal action of the Lie
algebra~$\sltR$ is by the first-order differential operators
  \begin{equation}\label{eq:6}
     \begin{aligned}  \tilde h\: \phi &\longmapsto -2x\phi ' - 2\lambda
     \phi \\  \tilde e\: \phi &\longmapsto  -\phi ' \\  \tilde f\: \phi
     &\longmapsto  \;\;\,x^2\phi ' + 2\lambda x\phi \\ \end{aligned} 
  \end{equation}
Simple calculus manipulations verify the identity~\eqref{eq:2}, which here
says that the ostensibly second-order differential operator on each side of
  \begin{equation}\label{eq:7}
     \frac 12 \tilde{h}^2 + \tilde{e}\tilde{f} + \tilde{f}\tilde{e} = \frac
     12\tilde{h}^2 + \tilde{h} + 2\tilde{f}\tilde{e} 
  \end{equation}
is the scalar operator that multiplies~$\phi $ by~$4\lambda ^2-2\lambda $.
 
\subsubsection*{The universal enveloping algebra} Let $\Usl$~be the
unital associative algebra generated by~$\sltR$ subject to the relation
generated by equating commutators in~$\Usl$ with the Lie bracket in~$\sltR$:
  \begin{equation}\label{eq:8}
     xy-yx=[x,y],\qquad x,y\in \sltR .
  \end{equation}
Any linear representation of the Lie algebra~$\sltR$ extends to a left module
for this \emph{universal enveloping algebra} ~$\Usl$.  A special case
of~\eqref{eq:8} is the relation
  \begin{equation}\label{eq:9}
     ef=fe+h
  \end{equation}
in~$\Usl$, where $e,f,h\in \sltR$ are the matrices~\eqref{eq:3}.
Substitute~\eqref{eq:9} into the left hand side of~\eqref{eq:2} to immediately
obtain the right hand side of~\eqref{eq:2}.  That's it!  This simple
manipulation in the abstract algebra~$\Usl$ proves that the
relation~\eqref{eq:2} holds in \emph{every} linear representation of~$\sltR$,
and so in particular immediately proves~\eqref{eq:5} and~\eqref{eq:7}.  Observe
that the expression $h^2/2 + ef + fe$ lies in the center of~$\Usl$, and
therefore necessarily acts as a scalar in every irreducible representation, as
we also saw in the three representations considered above.  (However, this
centrality is not the main point we are making here.)

\bigskip

 The framework we introduce for symmetry in quantum field theory enables
similar universal computations in an abstract quiche~$\sr$, and these
computations imply relations in every quantum field theory on which
$\sr$~acts.  Computations are made with defects in the topological field
theory~$\sr$.

  \subsection*{Motivation: Algebras}

We first remark that the word `symmetry' is usually used for a transformation
of a mathematical object that preserves structure, so in particular is
invertible.  But if, say, a finite group~$G$ acts linearly on a complex
vector space~$V$, then one can take linear combinations of the linear
operators corresponding to elements of~$G$ to endow~$V$ with the structure of
a module over the group algebra~$\GA$.  In this way one can talk about an
``algebra of symmetries'', though the algebra contains many non-units
(noninvertible elements).  As an extreme case, we might now call the action
of $0\in \GA$ a ``symmetry'' of~$V$.  Whether or not such ``noninvertible
symmetries'' merit this nomenclature, it is these algebras of symmetries
that motivate our definitions in field theory.

\subsubsection*{Abstract algebras of symmetries}Let $A$~be an algebra, and
for simplicity suppose that the ground field is~$\CC$.  For our motivational
exposition it suffices to assume that $A$~and the modules that follow are
finite dimensional.  We do not use a topology on~$A$.  Let $R$~be the right
\emph{regular} $A$-module, i.e., the vector space underlying the algebra~$A$
together with the right module structure $R\times A\to R$ given by
multiplication.  Then an abstract algebra of symmetries is the pair~$(A,R)$.

\subsubsection*{Concrete algebras of symmetries}Let $V$~be a complex vector
space.  The action of~$(A,R)$ on~$V$ is specified by the data of a
pair~$(L,\theta )$: $L$~is a left $A$-module and $\theta $~is an isomorphism
of vector spaces
  \begin{equation}\label{eq:10}
     \theta \:R\,\otimes \mstrut _AL\xrightarrow{\;\;\cong \;\;}V
  \end{equation}
Since $R$~is the right \emph{regular} module, the tensor product in the domain
of~\eqref{eq:10} is $A\otimes \mstrut _{A}L$, and there is a natural
isomorphism of this mixing of~$A$ and~$L$ with the vector space that underlies
the left module~$L$.  Passing from a module to its underlying vector space is
such a natural process that we do not usually bother to articulate the
isomorphism~\eqref{eq:10}.  But in the field theory context there is no
operation analogous to `underlying vector space'; \autoref{thm:4} below
suggests such an operation is not possible.  Elements of~$A=\End_A(R)$ act as
linear operators on~$V$ via the isomorphism~$\theta $.

\subsubsection*{A generalization}We can relax the condition that $R$~be the
\emph{regular} module and consider~$(A,R)$ with \emph{any} right module.
Then this very general notion of abstract symmetry allows many possibilities.
For example, if $A=\CC$ is the ground field as an algebra and $R$~is any
vector space, then an $(A,R)$-action on a vector space~$V$ decomposes~$V$ as
a tensor product $R\otimes L$ for some vector space~$L$.  For
general~$(A,R)$, the algebra~ $\End_A(R)$ acts on~$R\otimes \mstrut _{A}L$
and so too on~$V$ via the isomorphism~$\theta $.

  \subsection*{Quiches and their action on quantum field theories}

  \begin{figure}[ht]
  \centering
  \includegraphics[scale=.7]{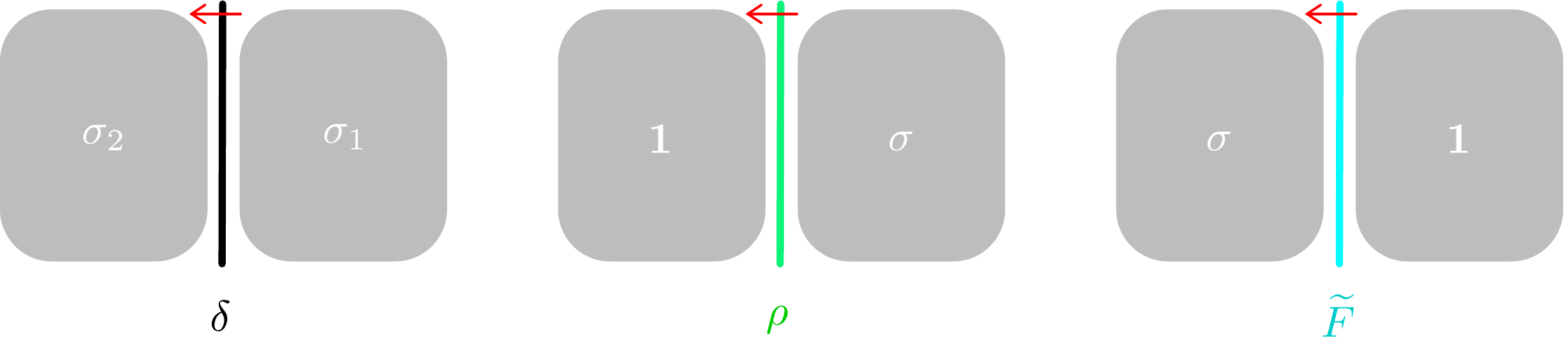}
  \vskip -.5pc
  \caption{A domain wall, a right boundary theory, and a left boundary theory}\label{fig:1}
  \end{figure}

\subsubsection*{Nomenclature: from algebra to field theory}
\begin{figure}[ht]
  \centering
  \includegraphics[scale=.8]{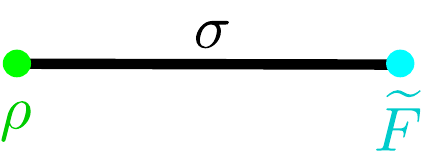}
  \vskip -.5pc
  \caption{The interval used in the dimensional
  reduction ~\eqref{eq:12}}\label{fig:2}
  \end{figure}

A Wick-rotated field
theory is a linear representation of a bordism category: this is the elevator
speech version of Graeme Segal's axiom system.  In this sense a Wick-rotated
field theory is analogous to a linear representation of a Lie group, or to a
left module over an algebra.\footnote{\label{alg}Suppose $A$~is an algebra and
$M$~is a right $A$-module.  Then one can contemplate an algebra~$A'$ that acts
on~$M$ on the left and commutes with the $A$-action.  For example, one often
considers the commutant $A'= \End _A(M)$.  In the analogy with field theory,
$A$~is the bordism category, the module~$M$ is the Wick-rotated field theory,
and $A'$~is the quiche, defined in the next few sections.}  We adopt
terminology from these analogs.  A field theory has a (spacetime) dimension and
a collection of background fields that determine the domain bordism category;
specifying these is analogous to specifying which Lie group or algebra we are
representing.  But in this note we often leave these important specifications
implicit.  Suppose $\sigma ,\sigma _1,\sigma _2$ are field theories (of the
same dimension on the same background fields).  We use the following
nomenclature for the domain walls and boundaries in \autoref{fig:1}:

  \begin{equation}\label{eq:11}
     \begin{aligned} \textnormal{domain wall $\delta \:\sigma _1\to\sigma
      _2$} &\;\mathbin{\squig\squig\squig\rsquigend}\; \textnormal{$(\sigma
      _2,\sigma _1)$-bimodule} \\ \textnormal{right boundary theory $\rho
      \:\sigma \to \bone$}\;\,&\;\mathbin{\squig\squig\squig\rsquigend}\;
      \textnormal{right $\sigma$-module} \\ \textnormal{left boundary theory\,
      $\tF \:\bone \to \sigma $}\;\,&\;\mathbin{\squig\squig\squig\rsquigend}\;
      \textnormal{left $\sigma$-module} \\ \end{aligned} 
  \end{equation}
Here $\bone$~is the tensor unit theory: $\bone\otimes F\cong F$ for all
theories~$F$.  Furthermore, we use the notation
  \begin{equation}\label{eq:12}
     \rho \otimes \mstrut _{\sigma }\tF 
  \end{equation}
for the dimensional reduction~$F$ of~$\sigma $ along the interval pictured in
\autoref{fig:2}, a presentation of~$F$ that we sometimes call a
\emph{sandwich}: the bulk theory~$\sigma $ is sandwiched between the right
boundary theory~$\rho $ and the left boundary theory~$\tF$.

\subsubsection*{Abstract field theoretic symmetry data: the quiche} The
nonnegative integer~$n$ in the following is the dimension of field theories
on which the quiche~$\sr$ acts.

  \begin{definition}[]\label{thm:2}
  Fix~$n\in \ZZ^{\ge0}$.  Then an \emph{$n$-dimensional quiche} is a
pair~$(\sigma ,\rho )$ in which $\sigma$~is an $(n+1)$-dimensional
{topological} field theory and $\rho $~is a right {topological} $\sigma
$-module.
  \end{definition}

\noindent
 It is the topological nature of the field theory~$\sigma $ and its right
boundary theory~$\rho $ that make this definition appropriate for topological
symmetries.  One can relax the definition and only require that $\sigma $~be
a once-categorified $n$-dimensional topological field theory; in its role as
symmetries of an $n$-dimensional field theory, it is not necessary to
evaluate~$\sigma $ on arbitrary $(n+1)$-manifolds.

  \begin{remark}[]\label{thm:7}
 In the motivational case of algebras, there is a distinguished right module:
the regular module.  Similarly, there are classes of topological field
theories~$\sigma $ for which we can define the notion of a regular $\sigma
$-module.  The examples in this note are of this type.
  \end{remark}

The most basic quiche is associated to a finite group~$G$ and any nonnegative
integer~$n$.  Let $\sigma $~be free $(n+1)$-dimensional gauge theory with gauge
group~$G$, and let $\rho $~be the right boundary theory which sums over
sections of the restriction of a principal $G$-bundle to the boundary.  (This
is sometimes called a Dirichlet boundary theory: the fluctuating principal
$G$-bundle is trivialized on the boundary.)  The detailed example at the end of
this note is a homotopical variant in which $G$~is the homotopical abelian
group~$BA$, the classifying space of a finite abelian group~$A$.

\subsubsection*{Action of a quiche on a quantum field theory}
\begin{figure}[ht]
  \centering
  \includegraphics[scale=.8]{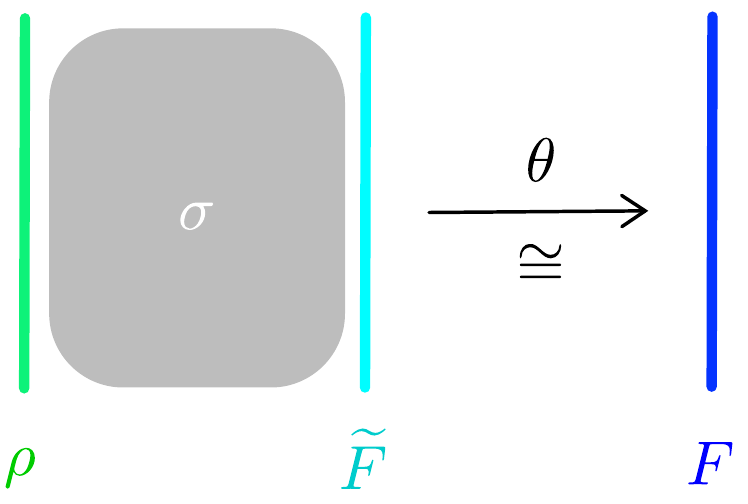}
  \vskip -.5pc
  \caption{$\sr$-module data on a field theory~$F$}\label{fig:3}
  \end{figure}

The following
definition is illustrated in \autoref{fig:3}.

  \begin{definition}[]\label{thm:3}
  Let $(\sigma ,\rho )$ be an $n$-dimensional quiche, and let $F$~be an
$n$-dimensional field theory.  A \emph{$\sr$-module structure} on~$F$ is a
pair~$\tFt$ in which $\tF$~is a left $\sigma $-module and $\theta $~is an
isomorphism
  \begin{equation}\label{eq:a21}
     \theta \:\rho \otimes \mstrut _{\sigma }\tF\xrightarrow{\;\;\cong \;\;}F 
  \end{equation}
of absolute $n$-dimensional theories.  
  \end{definition}

\noindent
 Neither~$F$ nor~$\tF$ is assumed to be topological.  (See
footnote~\footref{confess} for the mathematical status that we adopt in this
note and in~\cite{FMT} for nontopological field theories.)

  \begin{example}[center symmetry in gauge theory]\label{thm:4}
 Let $H$~be a Lie group and suppose $A\subset H$ is a finite subgroup of the
center of~$H$.  Then $A$~is abelian.  Set $\bH=H/A$.  (As an example, take
$H=\SU_2$ and $A=\bmut=\{\pm1\}\subset \SU_2$ the center; then $\bH\cong
\SO_3$.)  Suppose $F$~is an $H$-gauge theory in some dimension~$n$, and assume
that $F$~carries the homotopical symmetry of the group~$BA$, the classifying
space of~$A$.  For example, the theory may have matter fields that are neutral
under the central subgroup~$A\subset H$, in which case it has $BA$-symmetry.  A
map into~$BA$ classifies a principal $A$-bundle, and there is a bilinear
pairing $\{\textnormal{principal $A$-bundles}\}\times
\{\textnormal{$H$-connections}\}\to \{\textnormal{$H$-connections}$\}:
multiplication $A\times H\to H$ is a homomorphism of Lie groups since $A$~is
central.  We assume that the quantum field theory~$F$ has this action as a
symmetry.  (For example, this action does not alter the curvature, so pure
Yang-Mills theory has this symmetry.)  In the physics literature this is termed
a ``1-form symmetry''.  It is usually expressed by introducing a new background
field---an $A$-gerbe\footnote{An $A$-gerbe is a geometric representative of a
cohomology class in~$H^2(-;A)$; it arises here as the obstruction to lifting a
principal $\bH$-bundle to a principal $H$-bundle.  For $H=\SU_2$ and $A=\bmut$
this obstruction is the second Stiefel-Whitney class.}---and extending the
theory~$F$ to a theory with this \emph{background}/\emph{classical} field.
In~\cite{FMT} we advocate for the \emph{quantum} quiche picture: $\sigma $~ is
the quantum theory that sums over $A$-gerbes, $\rho $~is the boundary theory
that trivializes an $A$-gerbe, and $\tF$~is an $\bH$-gauge theory.  Observe
that no details about the quantum field theory~$F$ are used to describe the
symmetry; we only have to know that it carries a $BA$-symmetry.  In particular,
our characterization of~$F$ as ``an $H$-gauge theory'' is not necessary, and
furthermore this characterization only pertains to the description of the
theory, not to the abstract theory: the possibility of isomorphisms between
$H$-gauge theories and quantum field theories that lack a fluctuating
$H$-connection obstructs the statement ``The theory~$F$ is an $H$-gauge
theory''.
  \end{example}

  \subsection*{Quotients}

\subsubsection*{Augmentations of algebras}We begin as before with motivation
from algebra.

  \begin{definition}[]\label{thm:5}
 Let $A$~be an algebra over~$\CC$.  An \emph{augmentation} of~$A$ is an
algebra homomorphism $\eA \:A\to \CC$ from~$A$ to the ground field.
  \end{definition}

\noindent
 An augmentation gives the ground field~$\CC$ the structure of an $A$-module
$\CC_{\eA} $, which we take to be a right module: set $\lambda \cdot
a=\lambda\, \eA (a)$ for $\lambda \in \CC_{\eA} $, $a\in A$.  Let $R$~be the
right regular module, and let $V$~be a vector space equipped with an
$(A,R)$-action, including the structure isomorphism~\eqref{eq:10} that
recovers the underlying vector space~$V$.  \emph{Define} the vector space
  \begin{equation}\label{eq:13}
     Q:= \CC_{\eA} \otimes \mstrut _AL 
  \end{equation}
Then $Q$~is by definition the quotient of~$V$ by the action of~$(A,R)$ using
the augmentation~$\eA $.

  \begin{example}[]\label{thm:6}
 Let $G$~be a finite group and $A=\GA$ its complex group algebra.  A
character~$\chi $ of~$A$ determines the augmentation that sends $g\mapsto
\chi (g)$, $g\in G$.  In particular, there is a canonical augmentation from
the unit character: $\chi (g)=1$ for all~$g\in G$.  Let $S$~be a set with a
(left) $G$-action, and let $V=\CC\langle S \rangle$ be the free vector space
on~$S$.  Then for the unit character, $Q$~in~\eqref{eq:13} can be identified
with $\CC\langle S/G \rangle$, the free vector space on the quotient of~$S$
by the $G$-action.  Any character~$\chi $ determines a complex line bundle
$\sL\to S\gpd G$ over the groupoid (stack) quotient~$S\gpd G$.  In this case
$Q$~can be identified with the vector space of sections of $\sL\to S\gpd G$.
Note that at each~$s\in S$ the stabilizer subgroup $G_s\subset G$ acts on the
fiber~$\sL_s$ by a character.  If that character is nontrivial, then all
sections vanish at~$s$.
  \end{example}

Augmentations do not always exist and, as in \autoref{thm:6}, they may not be
unique.

\subsubsection*{Quotients in field theory}
\begin{figure}[ht]
  \centering
  \includegraphics[scale=.7]{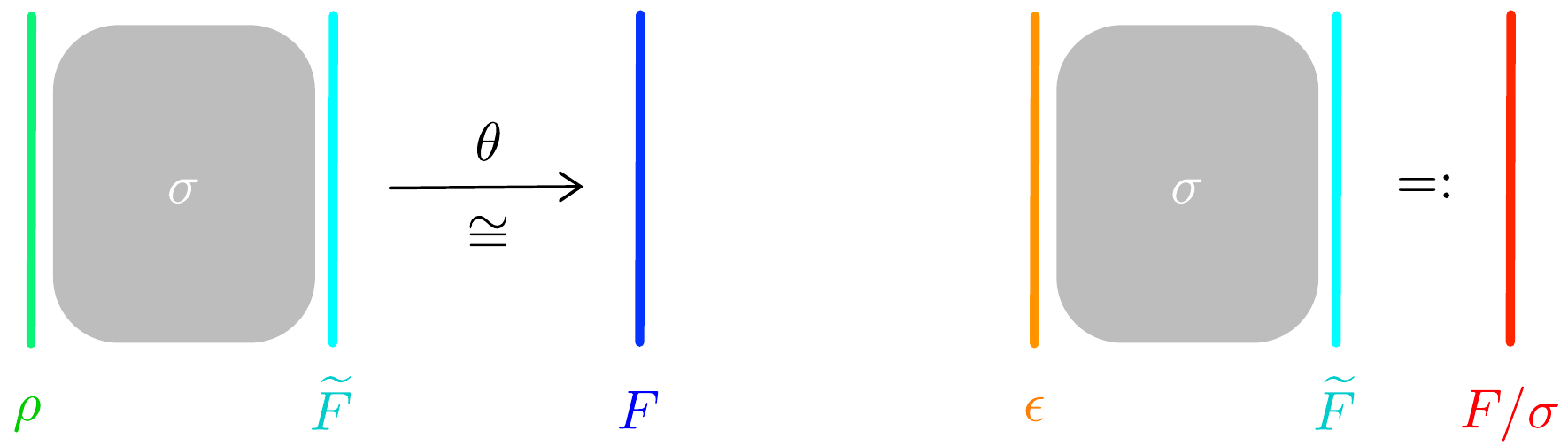}
  \vskip -.5pc
  \caption{\parbox[t]{30pc}{On the left, the structure isomorphism of the
  $\sr$-module structure on~$F$; on the right, the definition of the quotient
  theory~$F\bs$}}\label{fig:4}
  \end{figure}

This often goes under the name
`gauging a symmetry'.  Let $\sr$~be an $n$-dimensional quiche for some
nonnegative integer~$n$.  Just as the notion of a \emph{regular} module only
makes sense in special contexts (\autoref{thm:7}), so too does the notion of an
\emph{augmentation} of~$\sr$.  In fully local topological field theory, one
special context occurs when the codomain $(n+1)$-category of the theory~$\sigma
$ is the Morita category of algebras in some $n$-category.  In that case we
have the notions of both the regular module and augmentations: an augmentation
is an algebra map to the tensor unit.  We realize the augmentation~$\epsilon $
as a right $\sigma $-module.  Then if $\epsilon ^L$ is its left adjoint---a
particular dual left $\sigma $-module---the sandwich $\rho \otimes _\sigma
\epsilon ^L$ is the trivial theory.  Augmentations may not exist, whereas the
regular module of an algebra always exists.  For the gauge theory of a finite
group~$G$ introduced after \autoref{thm:7}, an augmentation is the right
boundary theory~$\epsilon $ which does not have any additional fluctuating
fields: there is no constraint on the bulk fluctuating $G$-bundle, hence the
quantization sums over all $G$-bundles.  As mentioned, this is the process of
quotienting, or gauging, (by) the $G$-symmetry.  The quotient theory along the
augmentation~$\epsilon $ is by definition
  \begin{equation}\label{eq:14}
     F\bs :=\epsilon \otimes\mstrut  _\sigma \tF,
  \end{equation}
as illustrated in \autoref{fig:4}.

  \subsection*{Line defects in 4-dimensional theories with $BA$-symmetry}

What follows is an extended example to illustrate the power of the {quantum}
quiche framework.  Our exposition is inspired by the papers~\cite{GMN,AST}.
We first state the result, which is a selection rule for line\footnote{Affine
geometry: point, line, plane.  Differential geometry: point, curve,
surface.  `Curve defects' anyone?} defects.  Then we briefly introduce
\emph{finite homotopy theories}, an important general class of topological
field theories.  We conclude with a derivation of the selection rule.

\subsubsection*{Setup and the selection rule}As in \autoref{thm:4}, let $H$~be
a Lie group and suppose $A\subset H$ is a finite subgroup of its center.
Suppose $F$~is a 4-dimensional field theory with $BA$-symmetry; we refer to it
as an $H$-gauge theory.  (But see the comment at the end of \autoref{thm:4}.)
For every pair~$(A',q)$ consisting of a subgroup~$A'\subset A$ and a quadratic
function $q\:A'\to \Cx$ there is an associated\footnote{Below we define the
theory~$F_{A',q}$ from~$F$ using only the $BA$-symmetry of~$F$;
see~\eqref{eq:34} and \autoref{fig:7}.  The descriptions as gauge theories are
not only superfluous but also do not have any intrinsic meaning.  Furthermore,
an important feature of our approach, as opposed to that in~\cite{GMN,AST}, is
that we give a \emph{local} construction of~$F_{A',q}$ from~$F$.
Characterizing a quantum field theory by cataloging its line defects is not a
local procedure.  Here we \emph{compute} the selection rule on line defects
from the local description of the theory, and moreover our computation takes
place entirely in the topological field theory that encodes the $BA$-symmetry.}
$H/A'$-gauge theory~$F_{A',q}$.  We compute a selection rule for line defects in
this theory.  As we explain below, the category of line defects is a sum of
categories indexed by
  \begin{equation}\label{eq:16}
     (m,e)\in A\times A\dual, 
  \end{equation}
where $A\dual=\Hom(A,\Cx)$ is the group of characters---the Pontrjagin dual
group---to~$A$.  Now the quadratic function~$q\:A'\to \Cx$ has an associated
bicharacter
  \begin{equation}\label{eq:30}
     b_q\:A'\times A'\to \Cx, 
  \end{equation}
and  so  by  transposition  a  homomorphism  
  \begin{equation}\label{eq:31}
     \epsilon _q\:A'\to (A')\dual. 
  \end{equation}
The selection rule for line defects in the theory~$F_{A',q}$ is:
  \begin{equation}\label{eq:17}
     \boxed{\begin{aligned} m&\in A' \\ e \res{A'}&=  \epsilon _q(m)\inv\end{aligned} }
  \end{equation}

For later use: The quadratic function~$q$ gives rise to a \emph{Pontrjagin
square} cohomology operation
  \begin{equation}\label{eq:15}
     \scaP_q\:H^2(X;A')\longrightarrow H^4(X;\Cx) 
  \end{equation}
on any topological space~$X$, and it too is a quadratic function:
  \begin{equation}\label{eq:33}
     \scaP_q(x_1+x_2) = \scaP_q(x_1) \cdot  \scaP_q(x_2) \cdot  \beta _q(x_1,
     x_2),\qquad x_1,x_2\in H^2(X;A'),
  \end{equation}
where $\beta _q\:H^2(X;A')\times H^2(X;A')\to H^4(X;\Cx)$ is the symmetric
bihomomorphism defined by the cup product and the bicharacter~\eqref{eq:30} on
coefficients.

\subsubsection*{Finite homotopy theories}

These topological field theories are constructed by a finite, homotopical
version of the Feynman path integral, hence are amenable to computations using
techniques from algebraic topology.  Some boundaries and defects in these
theories have semiclassical descriptions, so are similarly susceptible to
straightforward computation.  The data that defines a finite homotopy theory is
a pair~$(\sX,\lambda )$ consisting of a $\pi $-finite space\footnote{The $\pi
$-finiteness condition---$\sX$~has finitely many path components; there exists
$Q\in \ZZ^{\ge0}$ such that $\pi _q(\sX,x)=0$ for all $q>Q$, $x\in \sX$; and
$\pi _q(\sX,x)$~is a finite group for all~$q\in \ZZ^{\ge0}$, $x\in \sX$---is
only needed to define the partition function in the top dimension.  Without
that assumption we obtain a once-categorified field theory.  Our example
$\sX=\BtA$ \emph{is} $\pi $-finite, so defines a full topological field
theory.}~$\sX$ and a cocycle~$\lambda $ on~$\sX$; the theory exists in any
dimension.  In our example we use cocycles for singular cohomology with
coefficients in~$\Cx$.

Set $(\sX,\lambda )=(\BtA,1)$, where $\BtA$~is an Eilenberg-MacLane
space~$K(A,2)$ and $1$ is the identity cocycle that represents the identity
element in $H^4(\BtA;\Cx)$.  We construct a 5-dimensional topological field
theory~$\sigma $.  Let $M$~be a closed manifold of dimension~$\le4$; we briefly
describe the quantization~$\sigma (M)$ for $\dim M=4,3,2$.  Consider the
mapping space
  \begin{equation}\label{eq:18}
     \Xm M = \Map(M,\BtA). 
  \end{equation}
This mapping space is the space of ``fluctuating fields'' of the theory on~$M$;
its quantizations~\eqref{eq:20}--\eqref{eq:22} proceed via a homotopical form
of integration.  Eilenberg-MacLane spaces classify singular cohomology, from
which it follows that
  \begin{equation}\label{eq:19}
     \begin{aligned} \pi _0\bigl(\Xm M \bigr)&= H^2(M;A)
      \\ \pi _1\bigl(\Xm M \bigr)&= H^1(M;A) \\ \pi
      _2\bigl(\Xm M \bigr)&= H^0(M;A) \\ \end{aligned} 
  \end{equation}
Also, the mapping space $\Xm M=\Map(M,\BtA)$ is a product of
Eilenberg-MacLane spaces: all $k$-invariants vanish.  For $\dim M=4$ the
vector space~$\sigma (M)$ is the space of locally constant complex functions
on~$\Xm M$:
  \begin{equation}\label{eq:20}
     \sigma (M) = \Fun\flat(\Xm M)=\Fun(\pi _0\Xm M),\qquad \dim M=4. 
  \end{equation}
Notice that this quantization only uses the set~$\pi _0\Xm M$.  For $\dim
M=3$ the linear category~$\sigma (M)$ is the category of flat complex vector
bundles on~$\Xm M$:
  \begin{equation}\label{eq:21}
     \sigma (M) = \Vect\flat(\Xm M)=\Vect(\pi _{\le1}\Xm M),\qquad \dim M=3. 
  \end{equation}
Let $\VV$~denote the linear category of finite dimensional complex vector
spaces and linear maps.  Notice that the quantization~\eqref{eq:21} only uses
the fundamental groupoid~$\pi _{\le1}\Xm M$; the category~$\sigma (M)$ is
equivalent to the category of complex vector bundles over that groupoid, i.e.,
the category of functors $\pi _{\le1}\Xm M\to \VV$.  Now a \emph{$\VV$-module}
is a complex linear category, and a \emph{flat $\VV$-bundle} over a topological
space is a local system of $\VV$-modules.  For $\dim M=2$ the linear
2-category~$\sigma (M)$ is the 2-category of flat $\VV$-bundles on~$\Xm M$:
  \begin{equation}\label{eq:22}
     \sigma (M) = \Cat\flat(\Xm M)=\Cat(\pi _{\le2}\Xm M),\qquad \dim
     M=2. 
  \end{equation}
Let $\sC$~denote the linear 2-category of suitable complex linear categories.
Again the quantization~\eqref{eq:22} depends only on a truncation of the
mapping space, here the fundamental 2-groupoid $\pi _{\le2}\Xm M$, and
\eqref{eq:22}~is the 2-category of functors $\pi _{\le2}\Xm M\to \sC$.  Thus to
each object of $\pi _{\le2}\Xm M$ is assigned a linear category, to each
1-morphism is assigned a linear functor, and to each 2-morphism is assigned a
natural transformation of functors.  In particular, an elements of~$\pi _2(\Xm
M,\phi )$ acts as an automorphism of the identity functor of the category
attached to $\phi \in \Xm M$. 
 
There are semiclassical data that specify some special topological boundary
theories and topological defects in a finite homotopy theory.  The
semiclassical data of a topological boundary theory is a ($\pi $-finite) map
$f\:\sY\to \sX$ together with a suitable cochain on~$\sY$; for a topological
defect the codomain is an iterated free loop space of~$\sX$.  A regular
boundary theory has~$\sY=*$ a point; an augmentation has $f=\id\mstrut _{\sX}$.
We defer to~\cite[Appendix~A]{FMT} for details.

\subsubsection*{Proof of the selection rule}

Recall the setup in \autoref{thm:4}.  Namely, $A$~is a finite abelian group,
$\sigma $~is the 5-dimensional finite homotopy theory constructed
from~$(\BtA,1)$, and $\rho $~is the right regular boundary theory that
quantizes a basepoint $*\to \BtA$.  Furthermore, $F$~is a 4-dimensional
quantum field theory with $BA$-symmetry, and $(\tF,\theta )$ is the data that
expresses the $BA$-symmetry.  A special case already mentioned is gauge
theory: $H$~is a Lie group that contains~$A$ as a subgroup of its center,
then set $\bH=H/A$, and suppose~$F$ as an $H$-gauge theory with
$BA$-symmetry, in which case $\tF$~is an $\bH$-gauge theory (obtained by
coupling $H$-connections to $A$-gerbes).

\begin{figure}[ht]
  \centering
  \includegraphics[scale=.7]{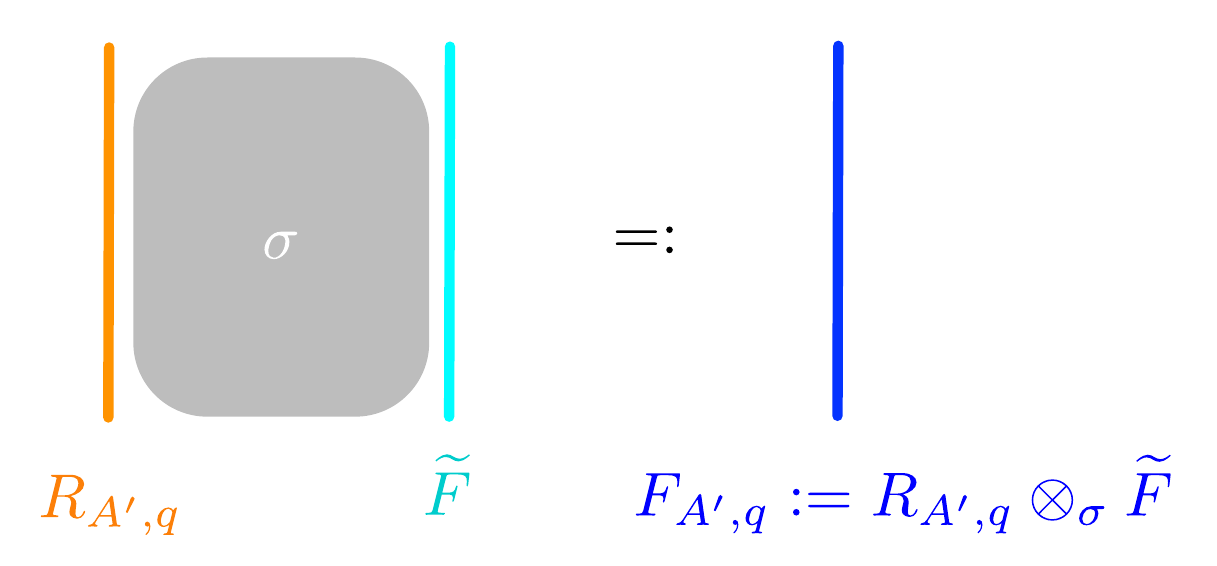}
  \vskip -.5pc
  \caption{The $q$-twisted quotient of~$F$ by $BA'$}\label{fig:7}
  \end{figure}

As stated in the text following~\eqref{eq:22}, semiclassical data for a
right boundary theory is a $\pi $-finite map $\sY\to \BtA$ together with a
cocycle~$\mu $ on~$\sY$ for a class in~$H^4(\sY;\Cx)$.  If $A'\subset A$ is a
subgroup, then $\sY=B^2\!A'\to \BtA$ is a $\pi $-finite map.  There is an
isomorphism 
  \begin{equation}\label{eq:23}
     H^4(\BnA2';\Cx) \cong \{\textnormal{quadratic functions
     }q\:A'\longrightarrow \Cx\} 
  \end{equation}
by a classical computation of Eilenberg-MacLane: the quadratic function~$q$
maps to the Pontrjagin square $\scaP_q(\iota )\in H^4(\BtAp;\Cx)$ of the
tautological class ~$\iota \in H^2(\BtAp;A')$.  Hence a pair~$(A',q)$
determines a right topological boundary theory~$\RAq$ for~$\sigma $.
Furthermore, the sandwich 
  \begin{equation}\label{eq:34}
     F_{A',q}:=\RAq\otimes \mstrut _{\sigma }\tF 
  \end{equation}
is the 4-dimensional quantum field theory which is a twisted quotient by~$BA'$,
as illustrated in \autoref{fig:7}.  In the language of gauge theories,
$\RAq\otimes \mstrut _{\sigma }\tF$ is a theory with gauge group~$H/A'$; the
quadratic form determines topological terms in the action of that gauge
theory. 

\begin{figure}[ht]
  \centering
  \includegraphics[scale=.9]{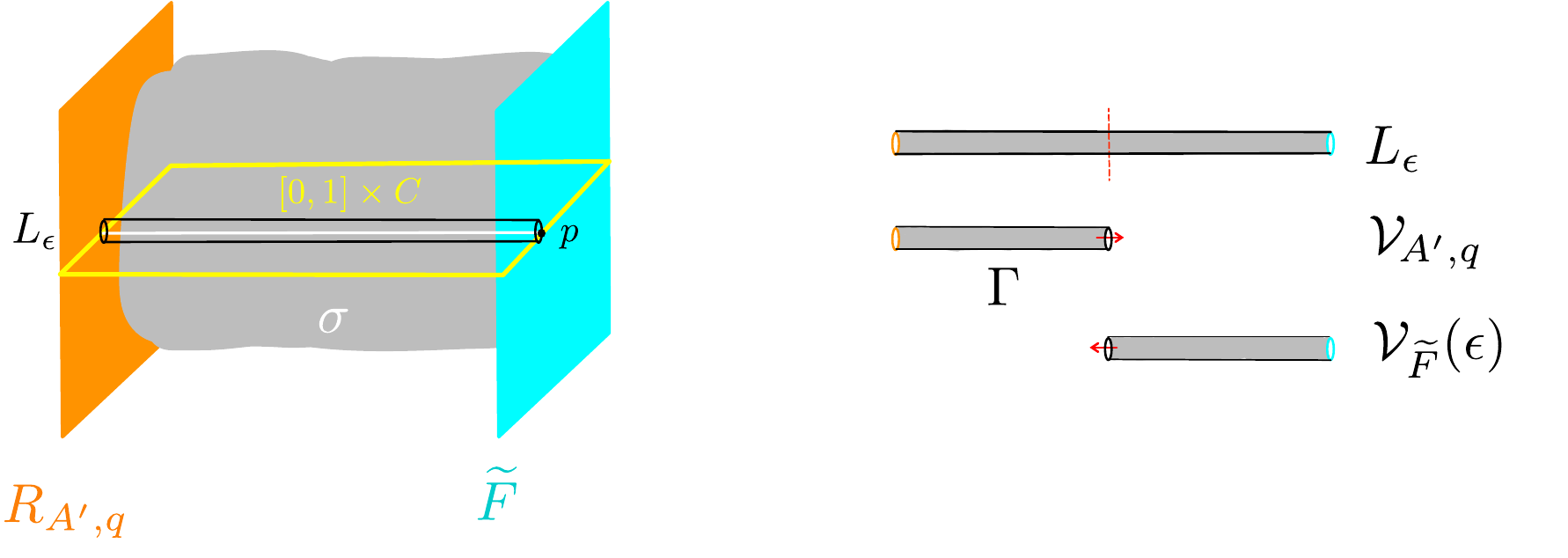}
  \vskip -.5pc
  \caption{The 1-category of line defects}\label{fig:10}
  \end{figure}

Fix~$\Aq$.  We investigate line defects in the theory $\RAq\otimes \mstrut
_{\sigma }\tF$.  The computation is local, but for ease of language assume
given a 4-manifold~ $M$~ and a 1-dimensional submanifold~ $C\subset M$ on which
we ``wrap'' defects.  In the sandwich picture, these are defects supported on
the 2-dimensional submanifold $[0,1]\times C\subset [0,1]\times M$; see
\autoref{fig:10}.  The 1-category of line defects\footnote{The global statement
on~$C$ is correct if the normal bundle to~$C\subset M$ is framed.  In any case,
it is the local statement at a point of~$C$ that is our focus here.} supported
on~$C\subset M$ is the inverse limit $\varprojlim\limits_{\epsilon \to
0}\Hom\bigl(1,\sigma (L_\epsilon ) \bigr)$, where in the sandwich picture
$L_\epsilon $~is the Cartesian product of~$[0,1]$ and the linking 2-sphere of
size~$\epsilon $ of $ C\subset M$ at some~$p\in C$.  To compute~$\sigma$ on
$L_\epsilon \approx [0,1]\times S^2$, cut along~$\{1/2\}\times S^2$ and so
express~$\sigma (L_\epsilon )$ as the 1-category
  \begin{equation}\label{eq:32}
     \Hom\bigl(\sV_{\Aq},\sV_{\tF}(\epsilon )\bigr) 
  \end{equation}
between objects in the 2-category~$\sigma (S^2)$.\footnote{To glue we must
reverse the depicted arrow of time at the boundary of the top
half-cylinder~$\Gamma $.  The finite semisimplicity of~$\sigma (S^2)$ makes
this a straightforward operation.}  Observe that $\sV_{\Aq}$~is
topological---the top half-cylinder~$\Gamma $ in \autoref{fig:10} lies in the
domain of the topological field theory~$(\sigma ,\RAq)$---so no
$\lim\limits_{\epsilon \to0}$ is required.  The limit is required
for~$\sV_{\tF}(\epsilon )$, but the derivation of the selection
rule~\eqref{eq:17} uses only the top half-cylinder~$\Gamma $, so we do not
encounter the limit in the sequel.

  \begin{remark}[]\label{thm:9}
 This slicing of~$L_\epsilon $ is the maneuver that sequesters the topological
computation from the nontopological part of the theory.  Such sequestration
maneuvers on defects are what the sandwich presentation of the theory make
possible. 
  \end{remark}

Next, we compute the 2-category~$\sigma (S^2)$ along which we factor
in~\eqref{eq:32}.  From~\eqref{eq:19} we have
  \begin{equation}\label{eq:24}
     \begin{aligned} \pi _0\bigl(\Map(S^2,\BnA2) \bigr)&= H^2(S^2;A)\cong A
      \\ \pi _1\bigl(\Map(S^2,\BnA2) \bigr)&= H^1(S^2;A)=0 \\ \pi
      _2\bigl(\Map(S^2,\BnA2) \bigr)&= H^0(S^2;A)\cong A \\ \end{aligned} 
  \end{equation}
Therefore, according to~\eqref{eq:22} the quantization~$\sigma (S^2)$ of~$\sM$
is the 2-category of flat $\VV$-bundles $\sV\to \pi _{\le2}\sM$.  The
computation~\eqref{eq:24} implies that for each $m\in \pi _0\sM$ we have a
linear category~$\sV^{(m)}$ equipped with an action of~$\pi _2\cong A$ by
automorphisms of the identity functor, hence $\sV^{(m)}$~decomposes as
  \begin{equation}\label{eq:25}
     \sV^{(m)} = \bigoplus\limits_{e}\sV^{(m,e)}\cdot e, \qquad e\in
     H^0(S^2;A)\dual\cong A\dual .
  \end{equation}
In the gauge theory context, $m$~and $e$~are called discrete magnetic and
electric fluxes.

As preparation for the final step of the computation, fix a cocycle~$\mu _q$
that represents the class in $H^4(\BtAp;\Cx)$ that corresponds to~$q$ under
the isomorphism~\eqref{eq:23}.  Suppose $f_1,f_2\:S^2\to \BtAp$ are maps
whose homotopy classes are $m_1,m_2\in A'$, respectively.  Then 
  \begin{equation}\label{eq:29}
     \bigl\langle \,(f_1\times f_2)^*(\mu _q)\,,\,[S^2\times
     S^2] \,\bigr\rangle = \beta _q(m_1,m_2),
  \end{equation}
follows from an application of~\eqref{eq:33} with $x_i=f_i^*(\iota )$,
where $\iota \in H^2(\BtAp;A')$ is the tautological class.

  \begin{figure}[ht]
  \centering
  \includegraphics[scale=.54]{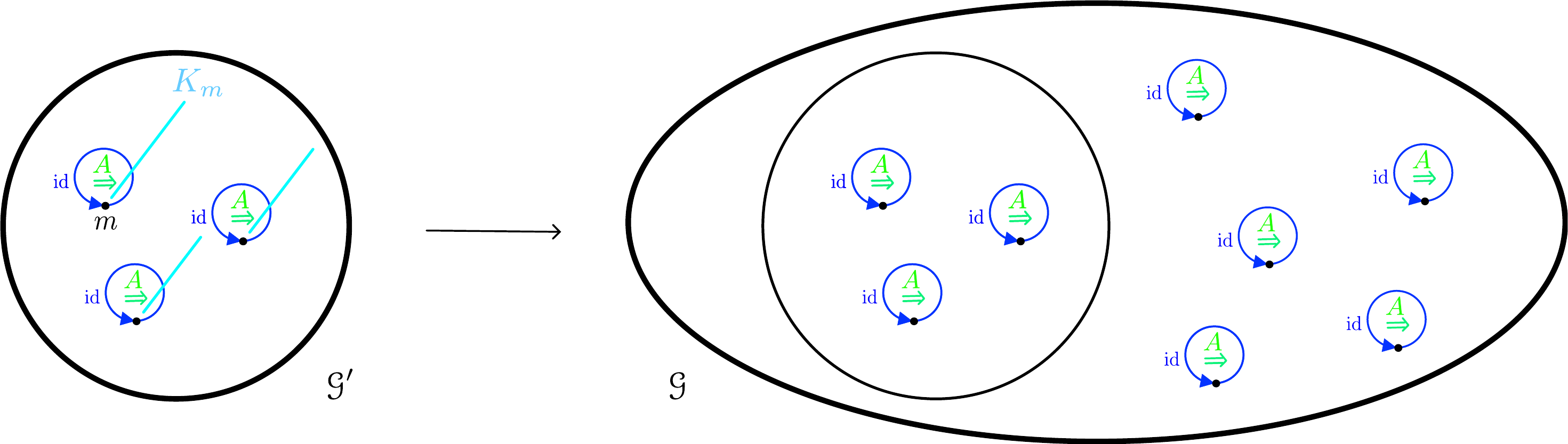}
  \vskip -.5pc
  \caption{The map $\sG'\longrightarrow \sG$ of 2-groupoids; see~\eqref{eq:28}}\label{fig:11}
  \end{figure}

Finally, we compute the value of~$(\sigma ,\RAq)$ on the half-cylinder~$\Gamma
$ in \autoref{fig:10}; the result is an object~$\sV\mstrut _{\Aq}=\bigoplus
_{m,e}\,\VAq^{(m,e)}\cdot e$ in the 2-category~$\sigma (S^2)$.  Note that $\Gamma $~
is a morphism $\emptyset ^2\to S^2$ in the bordism category; the
$\RAq$-coloring of the left boundary has the effect of ``coning off'' that
boundary component.  Intuitively, $\Gamma \:\emptyset ^2\to S^2$ is the bordism
that computes the value of the boundary theory~$\RAq$ on~$S^2$.  The
semiclassical value of this bordism is the correspondence
  \begin{equation}\label{eq:26}
     \begin{gathered} \xymatrix@R+1pc@C-2pc{&\bigl(\Map(S^2,B^2\!A'),\tau ^2(\mu
     _q)\bigr)\ar[dl]_{p_0} \ar[dr]^{p_1}\\ \ast&&
     \Map(S^2,\BtA)} \end{gathered}   
  \end{equation}
in which the cocycle $\tau ^2(\mu _q)$~is the transgression of~$\mu _q$; its
cohomology class lies in $H^2\bigl(\Map(S^2,B^2\!A');\Cx \bigr)$.
Represent~$\tau ^2(\mu _q)$ as a flat $\VV$-line bundle $\sK\to
\Map(S^2,B^2\!A')$.  Formally, the quantization of~$\Gamma $ is
  \begin{equation}\label{eq:27}
     \VAq = (p_1)_*\bigl(\sK\otimes p_0^*(\underline{\VV}) \bigr) =
     (p_1)_*(\sK),  
  \end{equation}
where $\underline{\VV}\to *$ is the trivial $\VV$-line bundle.  We compute
$(p_1)_*(\sK)$ using the map
  \begin{equation}\label{eq:28}
     \sG':=\pi _{\le2}\Map(S^2,B^2\!A')\xrightarrow{\;\;j \;\;} \pi
     _{\le2}\Map(S^2,\BtA)=:\sG  
  \end{equation}
of fundamental 2-groupoids, as depicted in \autoref{fig:11}.  (Recall the
homotopy groups~\eqref{eq:24} and the fact that all $k$-invariants vanish.
Hence we have strict models in which \eqref{eq:28}~ is an inclusion.)  Our
task is to compute the pushforward~$j _*(\sK)\to \sG$ of the $\VV$-line
bundle $\sK\to \sG'$.  First, it follows from~\eqref{eq:29} that for~$m\in
A'$, the action of $\pi _2(\sG',m)\cong A'$ on~$\sK_m$ is via the
character~$\epsilon _q(m)$, where recall the definition~\eqref{eq:31}
of~$\epsilon _q$.  Now since $j $~induces an inclusion $A'\hookrightarrow A$
on~$\pi _0$, it follows that $\VAq^{(m,e)}=0$ if~$m\notin A'$.  This proves
the first of the selection rules~\eqref{eq:17}.  For the second, observe that
for~$m\in A'$, the pushforward of~$\sK_m$ yields a $\VV$-module over~$\sG$
supported at $m\in A'\subset A$ on which the action of~$\pi _2(\sG,m)\cong A$
is by the induced representation $\Ind_{A'}^A\bigl(\epsilon _q(m) \bigr)$.
The latter is the sum of 1-dimensional representations: characters of~$A$
that restrict to~$\epsilon _q(m)$ on~$A'$ appear with multiplicity one.
Therefore, the $\VV$-bundle $\sV_{\Aq}\to \sG$ has support on~$A'\subset A$
and the fiber at~$m\in A'$ is a sum of $\VV$-lines indexed by characters
of~$A$ whose restriction to~$A'$ is $\epsilon _q(m)$.  It only remains to
observe that the inverse in the second selection rule in~\eqref{eq:17} is due
to the fact that $\VAq$~is in the domain of~\eqref{eq:32}, so appears
dualized.

 \bigskip\bigskip

\providecommand{\bysame}{\leavevmode\hbox to3em{\hrulefill}\thinspace}
\providecommand{\MR}{\relax\ifhmode\unskip\space\fi MR }
\providecommand{\MRhref}[2]{%
  \href{http://www.ams.org/mathscinet-getitem?mr=#1}{#2}
}
\providecommand{\href}[2]{#2}

\end{document}